\documentclass{aa}
\usepackage{graphicx}
\usepackage{alltt}
\usepackage{wasysym}
\usepackage{amssymb}  
\begin{document}

\def\hig{}
\def\half{\frac{1}{2}}

\title{The small and large lags of the elastic and anelastic tides. The virtual identity of two rheophysical theories.}

\author{Sylvio Ferraz-Mello}
\institute{Instituto de Astronomia 
Geof\'{\i}sica e Ci\^encias Atmosf\'ericas\\Universidade de S\~ao Paulo, Brasil \\
sylvio [at] usp.br}

\date{Received xx February 2015 / Accepted <date>}

\abstract{
The aim of this letter is to discuss the virtual identity of two recent tidal theories: our creep tide theory and one Maxwell model recently developed. It includes the discussion of the basic equations of the theories, which, in both cases, include an elastic and an anelastic component, and shows that the basic equations of the two theories are equivalent and differ by only a numerical factor in the anelastic tide.
It also includes a discussion of the lags: the lag of the full tide (geodetic), dominated by the elastic component, and the phase of the anelastic tide.  
In rotating rocky bodies  not trapped in a spin-orbit resonance (\textit{e.g.}, the Earth) the geodetic lag is close to zero and the phase of the semidiurnal argument in the anelastic tide is close to 90 degrees. The results obtained from
combining tidal solutions from satellite tracking data and from the Topex/Poseidon satellite altimeter data are extended to determine the phase of the semi-diurnal argument in the Earth's anelastic tide as $ \sigma_0=89.80\pm 0.05$ degrees. }
%
\titlerunning{Identity of two rheophysical theories}
\maketitle

\section{Introduction}

In his celebrated work on the secular change of elements due to tides, Darwin (1880) introduced {ad hoc} lags in the potential of a tidally deformed body. These lags were small and proportional to the frequencies of the delayed terms of the potential. Over the years, the fact that these two attributes were hypotheses introduced by Darwin has been forgotten. During the past century, some models with constant lags were attempted, but a serious discussion of the proportionality to frequency hypothesis had to wait for the work of Efroimsky and Lainey (2007) showing that in the case of terrestrial planets and planetary satellites, an inverse power law would be a better choice. But the lags continued to be small and the 
smallness of the tidal lags remained as a principle carved in stone. 

The contradiction between being small and following an inverse power law dominated discussions in the past decade and even served to argue the impossibility of the inverse power law. New results in contradiction with that postulate came from the creep tide theory (Ferraz-Mello, 2012, 2013).
One of the intermediary results of the creep tide theory is the shape of the body deformation due to the anelastic tide, which, after the transient phase (i.e. for $\gamma t \gg 1$), is dominated by the semidiurnal component. On the equatorial plane, the semidiurnal anelastic tide is
\begin{equation}\label{eq:semidi}
\zeta_{\rm an} =  \half R_e    \overline\epsilon_\rho E_{2,0}  
\cos  \sigma_0 \cos (2\varphi-2\ell- \sigma_0) 
\end{equation}
where
$  \overline\epsilon_\rho$ is the prolateness of the ideal Jeans spheroid representing the static tide due to an external body $\tens{M}$ at the distance $a$ (semi-major axis); $\varphi$ is the longitude of one point on the surface of the body; $R_e$ is the mean equatorial radius; $e$ is the orbital eccentricity; $\ell$ is the mean longitude of the external body; $E_{2,0}$ is the eccentricity function
\begin{displaymath}
E_{2,0}=1 - \frac{5}{2} e^2 + \cdots 
\end{displaymath}
and 
\begin{equation}\label{eq:sigma}
\tan \sigma_0 \equiv\frac{\nu}{\gamma}
\end{equation}
where $\gamma$ is the relaxation factor (also known as critical frequency) and $\nu=2\Omega-2n$ is the semidiurnal frequency ($\Omega$ is the rotation velocity of the body and $n$ is the orbital mean-motion). We remember that in the creep tide theory, $ \sigma_0$ is a fully determined constant introduced by the integration of the creep equation and not an {ad hoc} plugged lag.

The maximum of $\zeta_{\rm an}$ is reached when $2\varphi-2\ell- \sigma_0=0$, i.e. the angle between the vertex of the point where the height is maximum to the sub-$\tens{M}$ point is $ \sigma_0/2$. 
In the case of giant planets, the critical frequency $\gamma$ is in the range $10-100\ {\rm s}^{-1}$ (see Ferraz-Mello, 2013). 
The frequencies involved in the tide (rotation, mean-motion) are in the range 
$10^{-6} - 10^{-4}\ {\rm s}^{-1}$.  
Therefore, in this case $\nu \ll \gamma$ and so $ \sigma_0 \sim 0$ and the anelastic tide highest point remains almost aligned with the mean direction of the tide raising body $\tens{M}$.
However, in the case of rocky planets and planetary satellites, $\gamma$ is in the range $10^{-8}-10^{-7} \ {\rm s}^{-1}$. Then $\gamma\ll\nu$ and $ \sigma_0$ will approach 90 degrees. Besides, in intermediary cases approaching synchronization, $ \sigma_0$ may take any values in the interval $(0,\pi/2)$ and may lead to significantly large geodetic lags.

This last result contradicts the assumption of smallness of the lag postulated by Darwin, and some current beliefs. 
In the discussion of the creep tide theory, Ferraz-Mello (2013) stated that \textit {``this result is in contradiction with the observations. For instance, the observed geodetic lag of the Earth's body semidiurnal tide is very small} ($0.16 \pm 0.09$ degrees $cf.$ Ray et al. 1996)". This comment refers to the comparison of the geodetic lag with the lag of the anelastic creep tide and is not valid out of that context. In reality, as discussed in the next section, it is not in contradiction with the observations, but just with one particular interpretation of them. The value of $ \sigma_0$ close to 90 degrees is consistent with the observations and cannot be considered as a setback of the creep tide theory as stated by Correia et al. (2014).

\section{The observed value of the Earth's tidal lag}

We start this section quoting a statement found in Zschau (1978): 
\textit{``Measurements of tidal gravity variations at the Earth's surface, as well as precise observations of the tidal effect on satellite orbits have not yet revealed reliable results on imperfectly elastic body tides on Earth"}. In 2001, referring to earlier attempts to determine the energy dissipation in the Earth's body, Ray et al. (2001) said, \textit{``Unfortunately,  none of these early attempts to deduce $k_2$ from satellite tracking data was successful".} 
Only by the end of the century, combining tidal solutions from satellite tracking data and from Topex/Poseidon satellite altimeter data, Ray et al. (1996) managed to separate ocean and Earth tide signals and to determine the Earth body's dissipation. 
Their more recent result (Ray et al. 2001) corresponds to a geodetic lag of the Earth's body semidiurnal tide equal to  $\varepsilon_0 = 0.20 \pm 0.05$ degrees.

We recapitulate briefly how the lag was determined in the Ray et al. papers. They compare the Earth potential sensed by satellite tracking, which includes the contributions of the Earth's body, oceans and atmosphere, to the ocean component estimated from the Topex/Poseidon altimeter data. The difference is due to the other components which are accordingly modeled. The anelastic part of the tidal potential of the Earth's body is modeled using a classical Darwinian model. Using the notations of Ferraz-Mello et al. (2008), the potential component corresponding to the lunar semidiurnal tide, in the planar approximation, is
\begin{equation}\label{eq:u}
U_{2,0}=-\frac{k_f G m R^2}{5 r^{\ast 3}}  \overline\epsilon_\rho E_{2,0}
 \cos(2 \varphi-2\ell-\varepsilon_0)
\end{equation}
or, since $\varepsilon_0 \ll 1$, we obtain the anelastic contribution: 
\begin{equation}\label{eq:du}
\delta U_{2,0}=-\frac{k_f G m R^2}{5 r^{\ast 3}}  \overline\epsilon_\rho
E_{2,0} \sin(2 \varphi-2\ell)\sin\varepsilon_0
\end{equation}
where $k_f$ is the Love number for a homogeneous body and G the gravitational constant.
The static term (corresponding to $\varepsilon_0=0$) does not need to be considered as it was already taken into account in the two potential data being compared.
The anelastic part of the potential introduces $\sin\varepsilon_0$ in the Ray et al. equations, which, after elimination of the other effects (ocean loading and atmospheric tide and loading), is determined as $\varepsilon_0=0.20 \pm 0.05$ degrees.

However, we can proceed in a different way and use the creep tide theory instead of the Darwinian model. In that case, following Ferraz-Mello (2013), the component of the potential corresponding to the semidiurnal tide, in the planar approximation, is
\begin{equation}
\delta U_{2,0}=-\frac{2k_fGmR^2}{5r^{*3}}  \overline\epsilon_\rho E_{2,0}
\cos \sigma_0 \cos(2 \varphi-2\ell- \sigma_0).
\end{equation}
The angle $ \sigma_0$ is not an {ad hoc} lag, but a well-determined parameter that, in the case of the Earth, is very close to 90 degrees. This equation can then be expanded to become
\begin{equation}
\delta U_{2,0}=-\frac{2k_fGmR^2}{5r^{*3}}  \overline\epsilon_\rho E_{2,0}
\cos \sigma_0 \sin(2 \varphi-2\ell)\sin \sigma_0
\end{equation}
where we have neglected one term factored by $\cos^2  \sigma_0 \sim \gamma^2/\nu^2 \sim 0$. We have the same equations as before, but instead of $\sin\varepsilon_0$, we have
$\cos \sigma_0 \sin \sigma_0 = \frac{1}{2}\sin (\pi-2 \sigma_0) $. 
When this is used in the Ray et al. equations, the result is
$ \sigma_0=89.80 \pm 0.05$ degrees.\footnote{The comparison of the two results, valid when $\sigma_0\sim \pi/2$, gives $\varepsilon_0=\pi/2-\sigma_0$. }
The different approach for the introduction of the elastic tide in the creep tide theory does not affect the result because it was equally introduced in both tracking and altimeter data and only the difference between these two values actually matters in the determination of the lag.

\section{The equations of the creep tide theory}\label{sec:creep}

The creep tide theory of Ferraz-Mello (2013) is founded on one equation: the Newtonian creep differential equation
\begin{equation}\label{eq:7}
	\dot\zeta=-\gamma(\zeta-\rho)
\end{equation}
where $\zeta(\varphi,\theta)$ is the height of the anelastic tide\footnote{In Ferraz-Mello (2013) $\zeta$ was the distance to the center of the body and the height of the tides were represented by $\delta\zeta$ and $h$ (see Eqs. (55) and (58) in that paper). The change of the origin from the center of the body to one fixed reference level (e.g. a sphere with same volume as the body) simplifies notations when anelastic and elastic tides are composed and has no consequence in the study of the creep because both sides of Eq. (\ref{eq:7}) are invariant to such translation to the extent that $\rho$ is also referred to the same level as $\zeta$.} at one point at the surface of the body, $\rho(\varphi,\theta)$ is the {corresponding height of the} ellipsoidal figure of equilibrium due to the joint action of tide and rotation, and $\gamma$ is a relaxation factor inversely proportional to the {equivalent uniform} viscosity of the body. 
The solution is the anelastic tide, which is then added to the elastic tide $\lambda\rho$ (where $\lambda$ is a free parameter related to the height of the tide) to give the final result.

We can merge these two tide components into only one equation. 
If $Z=\zeta+\lambda\rho$, it is easy to see that 
\begin{equation}\label{eq:8}
	\dot{Z}+\gamma Z = (1+\lambda)\gamma\rho+ \lambda\dot{\rho}
\end{equation}
or,
\begin{equation}\label{eq:9}
	Z=Ce^{-\gamma t}+\lambda\rho + \gamma e^{-\gamma t}\int \rho e^{\gamma t} dt.
\end{equation}
{In order to see that Eq. (\ref{eq:9}) is the equation of a Maxwell body, it is enough to replace 
$\rho$ by the stress }
\begin{equation}
{\cal T}=\rho-\zeta=(1+\lambda)\rho-Z, 
\end{equation}
in the creep equation. Equation (\ref{eq:8}) then becomes
\begin{equation}
	\dot{Z} = (1+\lambda)\gamma {\cal T}+ \lambda\dot{\cal T},
\end{equation}
which is the constitutive equation of a Maxwell body (see Verh\'as, 1997).

\subsection{Comparison to the equations of the Maxwell model}
In order to compare the equations of the creep tide theory to those of the Maxwell model used by Correia et al. (2014), we need to rewrite Eq. (9) of Correia et al. using as a variable the deformation $\zeta$ instead of the potential $V_p$. Proceeding exactly as was done in that paper, but keeping $\zeta$ in the right-hand side of their Eq. (6), we obtain
\begin{equation}\label{eq:10}
	\dot{\zeta}+\frac{1}{\tau} \zeta = \frac{1}{\tau}\zeta_e + \frac{\tau_e}{\tau}\dot\zeta_e.
\end{equation}
If we substitute the two free parameters $\tau$ and $\tau_e$ by those used in the creep tide theory using  the equivalence formulas
\begin{equation}\label{eq:11}	
	\gamma=\frac{1}{\tau}, \qquad \qquad \lambda=\frac{\tau_e}{\tau},
\end{equation}
we obtain, for the basic equation of Correia et al. (2014),
\begin{equation}\label{eq:12}
	\dot{\zeta}+\gamma \zeta = \gamma\zeta_e+ \lambda\dot{\zeta_e}
\end{equation}
whose solution is
\begin{equation}\label{eq:13}
\zeta=Ce^{-\gamma t}+\lambda\zeta_e + \gamma (1-\lambda) e^{-\gamma t}\int \rho e^{\gamma t} dt.
\end{equation}
Before comparing this result to that of the creep tide model, we have to discuss the meaning of $\zeta$ in both theories. In Correia et al. (2014), $\zeta$ is the radial deformation of the free surface. In Ferraz-Mello (2013), $\zeta$ is also the deformation of the free surface, but before the inclusion of the elastic part in the theory. So, the $\zeta$ of the Maxwell model is to be compared not to $\zeta$ of the creep equation, but to $Z$.
We also note that $\zeta_e\equiv\rho$.

The comparison of Eqs. (\ref{eq:9}) and (\ref{eq:13}) shows that the two theories are virtually identical. The only difference between the two theories is the numerical factor $(1-\lambda)$, which appears in Eq. (\ref{eq:13}) multiplying the part of the solution corresponding to the anelastic tide. (We note that $0<\lambda<1$). 
This explains why the results of Correia et al. (2014) for the dissipation and the rotation look identical to those of Ferraz-Mello (2013, 2015). 

As an example, the empirical formulas relating the creep theory parameters to the dissipation parameter $k_2/Q$ of the classical theories are
\begin{displaymath}
\frac{k_2}{Q}=k_f\frac{\gamma\nu}{\gamma^2+\nu^2}
\end{displaymath}
in Ferraz-Mello et al. (2013)
and
\begin{displaymath}
\frac{k_2}{Q}={k_f}(1-\lambda)\frac{\gamma\nu}{\gamma^2+\nu^2}
\end{displaymath}
in Correia et al. (2014). {The equivalence between these two equations was already shown by Eq. (90) in Correia et al. (2014)}. 

\subsection{The Newtonian creep of the Maxwell model}

The inverse of the transformation used in Section \ref{sec:creep} can be applied to Eq. (\ref{eq:12}) to give the equation of the creep embedded in the Maxwell model of Correia et al. {If we denote by $\zeta_{\rm an}$ the radial deformation of the surface due to the anelastic component, that transformation is}
\begin{equation}
\zeta=\zeta_{\rm an}+\lambda\zeta_e
\end{equation} 
and the result is
\begin{equation}\label{eq:14}
\dot{\zeta_{\rm an}} = -\gamma \big(\zeta_{\rm an}-(1-\lambda)\zeta_e\big).
\end{equation}
This is a Newtonian creep which differs from that considered in Eq. (\ref{eq:7}) only by the fact that here the stress is taken proportional to the distance to a different equilibrium surface, defined by  $(1-\lambda)\rho$, instead of the surface of a Jeans ellipsoid, $\rho$, as in Eq. (\ref{eq:7}).

{The solution of Eq. (\ref{eq:14}) is trivial. It gives, for the semidiurnal anelastic tide of the Maxwell model,}
\begin{equation}\label{eq:15}
\zeta_{\rm an} =  \half R_e    \overline\epsilon_\rho E_{2,0}  
(1-\lambda)\cos  \sigma_0 \cos (2\varphi-2\ell- \sigma_0)
\end{equation}
{where $\sigma_0$ is the same angle defined by Eq. (\ref{eq:sigma}). This means that the phase of the anelastic tide in the Maxwell model of Correia et al. (2014) is the same as that of the creep tide theory of Ferraz-Mello (2013).}

\section{Conclusions and summary}

1. The dynamical tide can be decomposed into two parts: one elastic part, which corresponds to the perfect deformation of the body under the tidal stress, and the so-called anelastic part, which corresponds to having the body permanently adjusting its shape to follow the changing tidal potential. In the two theories discussed in this letter, Ferraz-Mello(2012, 2013) and Correia et al. (2014), these two parts are virtually the same. The only difference between them is that the solution given by Correia et al. for the anelastic tide is the same solution found in the creep tide theory of Ferraz-Mello multiplied by the numerical factor $(1-\tau_e/\tau) \equiv (1-\lambda)$.

The virtual identity of the two theories, notwithstanding their completely different formulations, can be considered as a source of insight for the understanding of the physical problem. The fact that the results can be obtained with the much simpler creep tide {model, before the introduction of the elastic tide,} is also insightful.

2. Two angles play a major role in tide theory, the geodetic lag ($\varepsilon_0$), which measures the asymmetry of the shape of the body, and the {phase} of the {semidiurnal argument in the} anelastic tide ($\sigma_0$), which measures the asymmetry of the shape of the anelastic deformation. In rotating bodies  not trapped in a spin-orbit resonance (like the Earth), with low relaxation factor, the geodetic lag is close to zero and the {phase} of the semidiurnal argument in the anelastic tide is close to 90 degrees. This result is the same in both theories: the creep tide theory of Ferraz-Mello (2013) and the Maxwell model of Correia et al. (2014). It does not depend on the used theory. It is also true for modern versions of Darwin's theory (Efroimsky, 2012a).

Bodies trapped in a spin-orbit resonance were not considered in this paper since, in that case, $\nu \rightarrow 0$ and then the {phase} of the anelastic tide ($\sigma_0$) tends to zero. This is also so in the frame of Efroimsky's theories (Efroimsky, 2012b; Williams and Efroimsky, 2012). In the particular resonant case of Mercury, the equations are somewhat different because the dominant tide has a different argument ($\nu-n$ instead of $\nu$; see Ferraz-Mello 2015). The anelastic tide is then dominated by the term whose phase is
\begin{displaymath}
\sigma_{-1}=\arctan \frac{\nu-n}{\gamma} \rightarrow 0.
\end{displaymath}

3. The Earth's lag determined by Ray et al. (2001) from the combination of tidal solutions from satellite tracking data and from Topex/Poseidon satellite altimeter data is the geodetic lag. 
The same data may be used to determine the {phase} of the {semidiurnal argument in the} anelastic tide and gives $\sigma_0=89.80\pm 0.05$ degrees.

4. The results of the Maxwell model of Correia et al. (2014) can be obtained in a much simpler way with the methodology used in Ferraz-Mello (2013). The main results can be obtained from the solution of the Maxwell model's creep (Eq. \ref{eq:14}) and a few elementary physics laws. The elastic component of the tide and its effects can be added afterwards.

\begin{acknowledgement}
This investigation is funded by the National Research Council, CNPq, grant 306146/2010-0. I thank Drs. Michael Efroimsky and Alexandre C.M. Correia for their valuable comments and suggestions.

\end{acknowledgement}

\end{document}